\documentclass[aps,twocolumn,prd,preprintnumbers, 10pt]{revtex4-1}
\pdfoutput=1
\usepackage{amsmath}
\usepackage{graphicx}
\usepackage{bbm}
\usepackage{amssymb}
\usepackage{mathrsfs}
\usepackage{slashed}
\usepackage{cancel}
\usepackage[linktocpage=true,
colorlinks=true,
urlcolor=blue,
anchorcolor=blue,
citecolor=blue,
filecolor=blue,
linkcolor=blue,
menucolor=blue
]{hyperref}
\usepackage[normalem]{ulem}
\usepackage{soul,xcolor}

\newcommand{\ex}{\mathrm{e}}
\newcommand{\diff}{\mathrm{d}}
\newcommand{\R}{\mathbb{R}}
\newcommand{\vol}{\mathrm{vol}}

\newcommand{\Fgrav}{F_{\mathrm{grav}}}

\newcommand\cA{\mathcal{A}}

\newcommand\cK{\mathcal{K}}
\newcommand\cG{\mathcal{G}}
\newcommand\cF{\mathcal{F}}
\newcommand\cN{\mathcal{N}}
\newcommand\cR{\mathcal{R}}
\newcommand\cV{\mathcal{V}}

\newcommand\cI{\mathcal{I}}
\newcommand{\e}{\mathrm{e}}
\newcommand{\ii}{\mathrm{i}}
\newcommand{\dd}{\mathrm{d}}

\newcommand{\tz}{\widetilde{z}}
\newcommand{\tepsilon}{\widetilde{\epsilon}}
\newcommand{\hook}{\mathbin{\rule[.2ex]{.4em}{.03em}\rule[.2ex]{.03em}{.9ex}}}

\newcommand{\abs}[1]{\left\lvert #1 \right\rvert}

\usepackage{color}

\def\nn{\nonumber}

\newcommand{\Z}{\mathbb{Z}}

\newcommand{\newP}{\mathbb{P}}
\newcommand{\pmu}{{\kappa}}

\newcommand{\xinew}{\zeta}

\newcommand{\selfd}{[+]}
\newcommand{\aselfd}{[-]}

\begin{document}

\setstcolor{red}

\title{Localization of the free energy in supergravity}

\author{Pietro Benetti Genolini$^{1}$}
\author{Jerome P. Gauntlett$^{2}$}
\author{Yusheng Jiao$^{2}$}
\author{Alice L\"uscher$^{3}$}
\author{James Sparks$^{3}$}

\affiliation{$^{1}$D\'epartment de Physique Th\'eorique, Universit\'e de Gen\`eve, 24 quai Ernest-Ansermet, 1211 Gen\`eve, Suisse}
\affiliation{$^{2}$Blackett Laboratory, Imperial College, Prince Consort Road, London, SW7 2AZ, U.K.}
\affiliation{$^{3}$Mathematical Institute, University of Oxford, Woodstock Road, Oxford, OX2 6GG, U.K.}

\begin{abstract}
\noindent  
We derive a  general formula for the  gravitational free energy  
of Euclidean supersymmetric solutions to $D=4$, $\mathcal{N}=2$ 
gauged supergravity coupled to vector multiplet matter. 
This allows one to compute the free energy
without solving 
any supergravity equations, just assuming the solutions exist. 
As well as recovering some known results in the literature with
ease, 
we also present new supergravity results that match with holographically dual 
field theory computations.

\end{abstract}

\maketitle

\section{Introduction}\label{sec:intro}
Supersymmetric supergravity solutions continue to play a key role in string/M-theory.
Such solutions solve the equations of motion and specific Killing spinor equations, which together comprise a 
challenging system of non-linear 
partial differential equations.  Many explicit solutions have been constructed over the years, using a variety of techniques, and have led to important insights in diverse settings. However, generically such constructions comprise a small subset of the solutions of physical interest and more general methods are highly desirable. It has recently been shown \cite{BenettiGenolini:2023kxp} that supersymmetric solutions with an R-symmetry possess a remarkable general structure whereby one can compute various BPS physical observables without solving the supergravity equations at all, just assuming that the solution exists and inputting some topological information. 

It is well-known that the Killing spinor equations 
may be 
recast as equations for differential 
forms constructed as Killing spinor bilinears (for an early paper, 
see \cite{Gauntlett:2002sc}). 
In many contexts this includes a Killing vector bilinear which rotates the Killing spinor, called the 
 \emph{R-symmetry vector $\xi$}.  It is then natural to introduce 
the equivariant exterior derivative $\dd_\xi \equiv \dd  - \xi \hook\, $, 
acting on polyforms constructed from the bilinear forms and supergravity fields.
In varied settings \cite{BenettiGenolini:2023kxp, BenettiGenolini:2023yfe, BenettiGenolini:2023ndb, BenettiGenolini:2024kyy, Suh:2024asy,Couzens:2024vbn} 
it has been shown that one can construct a number of 
$\dd_\xi$-closed 
such polyforms in a given supergravity theory. Moreover, the integrals of these forms give rise to 
various BPS physical observables, 
typically allowing one to compute such quantities just
using the BVAB
fixed point formula~\cite{BV:1982, Atiyah:1984px}. 
This constitutes a powerful new technique for computing observables in
supergravity, which underlies and greatly extends many results in the literature.

In this paper we derive a general formula for the on-shell action, 
or \emph{gravitational free energy}, of Euclidean supersymmetric solutions to $D=4$, $\mathcal{N}=2$ 
gauged supergravity \footnote{We focus on gauged supergravity, but our formalism immediately gives several results valid in ungauged supergravity too.}  coupled to vector multiplets. This significantly extends the results for minimal gauged supergravity in \cite{BenettiGenolini:2019jdz,BenettiGenolini:2023kxp}. The theory 
involves a prepotential $\cF=\cF(X^I)$, 
which is a holomorphic, homogeneous degree two function of the scalar fields $X^I$
in the vector multiplets. 
The R-symmetry vector $\xi$ for a supersymmetric solution
will in general have a fixed point set $\{\xi=0\}\subset M_4$ in the Euclidean spacetime 
$M_4$
 consisting of 
isolated fixed points (called \emph{nuts}) and/or fixed 
surfaces (called \emph{bolts}) \cite{Gibbons:1979xm}. 
As in \cite{BenettiGenolini:2019jdz, BenettiGenolini:2023kxp} 
we find that the Killing spinor $\epsilon$ is necessarily chiral  
at a fixed point, so $\gamma_5\epsilon=\pm \epsilon$, 
and we may 
correspondingly append this chirality 
data to the fixed point sets, 
 labelling them as
$\mathrm{nut}_\pm$, $\mathrm{bolt}_\pm$.  

The gravitational free energy only receives contributions from these fixed points and 
assuming, for simplicity, a reality condition on the scalars
and that $\ii\cF$ is real \footnote{These assumptions can be dropped leading to a more general formula.
}, we have
\begin{align}\label{Fgrav}
& \Fgrav = \frac{\pi}{G_4}\Bigg[\sum_{\mathrm{nuts}_\pm}\mp\frac{1}{d_\pm} \frac{(b_1\mp b_2)^2}{b_1b_2}\ii \cF(u^J_\pm) \nonumber
\\
& + \sum_{\mathrm{bolts}_\pm}\bigg( -\pmu \mskip2mu \partial_I \ii\cF(u^J_\pm) \mathfrak{p}^I_\pm\pm \ii \cF(u^J_\pm) \int_{\Sigma_\pm} c_1(L) \bigg)\Bigg]\, .
\end{align}
Here $G_4$ is the Newton constant and $\ii\cF$ 
is evaluated 
on the following combinations of scalars at the fixed points
\begin{align}\label{uI}
\left.u^I_+ \equiv \frac{\widetilde{X}^I}{\xinew_J \widetilde{X}^J}\right|_+\, , \quad 
\left.u^I_- \equiv \frac{{X}^I}{\xinew_J {X}^J}\right|_-\, .
\end{align}
The subscript $|_\pm$ denotes restriction to fixed points
of the  above-mentioned chirality, where
$u^I_\pm$ are necessarily constant 
over a connected component of the fixed point set. 
The $\xinew_I\in\R$ are Fayet--Iliopoulos (FI) gauging parameters of the theory, the complex conjugate of the scalar 
$X^I$
becomes an independent field in Euclidean signature, i.e. $\overline{X}^I\mapsto \widetilde{X}^I$,
and to obtain \eqref{Fgrav} we have taken $X^I,\widetilde{X}^I$ to be real.
Notice from \eqref{uI} that it is either $X^I$ or $\widetilde{X}^I$ 
appearing in the formula, correlated with the chirality. 

At a nut we have written  $\xi=\sum_{i=1}^2 b_i 
\partial_{\varphi_i}$, where $\partial_{\varphi_i}$ rotate each 
of $\R^2_i$ in the tangent space $\R^4=\R^2_1\oplus \R^2_2$ at the 
nut. 
The $b_i$ are hence the {weights} of $\xi$ at an isolated fixed point. The formula \eqref{Fgrav} also allows for orbifold singularities, 
where the tangent space is  $\R^4/\Gamma_{d_\pm}$ with $d_\pm$ the order of the  finite group  $\Gamma_{d_\pm}$.
We denote 
a fixed bolt by $\Sigma_\pm$, with 
$\mathfrak{p}^I_\pm\equiv 1/(4\pi)\int_{\Sigma_\pm}F^I$
the magnetic flux through $\Sigma_\pm$, where $F^I$ is the gauge field
curvature for the $I^{\mathrm{th}}$ vector multiplet, and 
$c_1(L)$ is the first Chern class of the normal bundle $L$ of 
$\Sigma_\pm\hookrightarrow M_4$. Supersymmetry
fixes the following R-symmetry flux through a bolt 
\begin{align}
 \label{RFlux}
 \mathfrak{p}_R\equiv \frac{1}{2}\zeta_I \mathfrak{p}^I_{\pm} = \frac{\pmu}{2}\int_{\Sigma_\pm} \left[\pm c_1(L) - c_1(T\Sigma_\pm) \right] \, .
\end{align} 
 If the bolt $\Sigma$ is a genus $g$ Riemann surface we have $\int_{\Sigma} c_1(T\Sigma)=\chi=2(1-g)$, the Euler number of $\Sigma$. 
Finally, $\pmu=\pm1$ in \eqref{Fgrav}, \eqref{RFlux}  is a sign for a given fixed point.

In \eqref{Fgrav}, $ \Fgrav$ is written
as a function of data at the fixed points of the R-symmetry, which 
are assumed to be in the interior of the Euclidean spacetime $M_4$. 
As we show, localization also allows one 
to interpret this as a function of data defined on the conformal 
boundary $M_3=\partial M_4$, which includes specifying the choice of (conformal) 
Killing vector $\xi$ on the boundary \footnote{Specifying the boundary Killing vector then determines the weights $b_i$ at the fixed points
in \eqref{Fgrav}.}, 
thus making contact with 
the free energy in a holographically dual CFT  defined on $M_3$.  
As well as recovering formulae for various 
supergravity solutions, knowing only their topology, we also
derive new results for which explicit solutions have not been constructed.

 \section{Euclidean theory and equivariant forms}
The Euclidean supergravity theory is obtained from Lorentzian $\mathcal{N}=2$ gauged supergravity coupled to $n$ vector multiplets
using the procedure in \cite{Freedman:2013oja}. The bosonic content of the Lorentzian theory consists of
a metric, $n+1$ gauge fields, $A^I$, $I=0,1,\dots n$, with field strengths $F^I=\dd A^I$,
as well as $n$ complex scalar fields $z^i$, $i=1,\dots, n$. The
$z^i$ parametrize a K\"ahler manifold with K\"ahler potential $\cK$ and Hermitian metric 
$\cG_{i\bar{j}}\equiv \partial_i\partial_{\bar j}\cK$; it is also the base of a symplectic bundle with covariantly constant sections
$\e^{\cK/2} (X^I , \cF_I)$, symplectic constraint $\e^{\cK}( X^I \overline{\cF}_I - \cF_I \overline{X}^I ) = \ii$, and where 
$\cF_I = {\partial_I \cF}$ is the derivative of the prepotential. 
Defining $\cN_{IJ} = \overline{\cF}_{IJ} + \ii \frac{N_{IK}X^K N_{JL}X^L}{N_{NM} X^N X^M}$,
where $\cF_{IJ} = \partial_I \partial_J \cF$, $N_{IJ} = 2\, \text{Im}\,   \cF_{IJ}$, we define $\cR_{IJ} \equiv \text{Re}\,  \cN_{IJ}$ and $\cI_{IJ} \equiv \text{Im}\, \cN_{IJ}$. Also note $\mathcal{N}_{IJ}X^J=\mathcal{F}_I$.

In the Euclidean theory, the metric and gauge fields are complex and we also have $2n$ complex scalar fields $z^i$, $\tz^{\tilde i}$. The  
(bulk) bosonic action is given by
\begin{align}\label{eq:I4dEucl}
I	=&- \frac{1}{16\pi G_4} \int \Big[ \big( R - 2 \cG_{i\tilde{j}} \partial^\mu z^i \partial_\mu \tz^{\tilde{j}} - \cV (z, \tz)  \big)\vol_4 \nn \\ 
	& + \frac{1}{2} {\cI}_{IJ} F^I \wedge * F^{J} - \frac{\ii}{2} {\cR}_{IJ} F^I \wedge F^J \Big] \, .
\end{align}
Here $\vol_4$ denotes the Riemannian volume form, with Hodge 
duality operator $*$, while 
 the potential is
	$\cV = \e^{\cK} \big( \cG^{i\tilde{j}} \nabla_i W \nabla_{\tilde{j}} \widetilde{W} - 3 W \widetilde{W} \big)$,
where
$W \equiv \xinew_I X^I$, $\widetilde W \equiv \xinew_I \widetilde X^I$, $\nabla_i X^I \equiv ( \partial_i + \partial_i \cK) X^I$ and the FI parameters $\zeta_I\in \mathbb{R}$, as in the Lorentzian theory.  
The Killing spinor equations, solved by supersymmetric configurations, 
can be parametrized by two Dirac spinors, $\epsilon$, $\tepsilon$ and are presented in the supplementary material. The Lorentzian theory is recovered
by taking the metric and gauge fields to be real, $\tz^{\tilde i}=\bar z^{\bar i}$ (and hence $\widetilde X^I=\overline X^I$, $\widetilde W=\overline W$) as well as
$\tepsilon=\epsilon^c$, where $\epsilon^c$ is the charge conjugate spinor.

For a supersymmetric solution we can construct the spinor bilinears
\begin{align}
\label{eq:Bilinears}
	& S \equiv \overline{\tepsilon} \epsilon \, ,  \quad &&P \equiv \overline{\tepsilon} \gamma_5 \epsilon \, ,  \quad 
	&&&\xi^\flat \equiv - \ii \mskip1mu \overline{\tepsilon} \gamma_{(1)} \gamma_5 \epsilon \, ,\nonumber\\
	& K \equiv \overline{\tepsilon} \gamma_{(1)} \epsilon \, , 	\quad 
		&& U \equiv \ii \mskip1mu \overline{\tepsilon} \gamma_{(2)} \epsilon \, ,
\end{align}
where $\overline{\epsilon} \equiv \epsilon^{\mathrm{T}}\mathcal{C}$ is the Majorana
 conjugate, the charge conjugation matrix $\mathcal{C}$ satisfies 
$\gamma_\mu^{\mathrm{T}}=\mathcal{C}\gamma_\mu\mathcal{C}^{-1}$, 
and we have defined $\gamma_{(r)}\equiv \frac{1}{r!}\gamma_{\mu_1\cdots \mu_r}
\diff x^{\mu_1}\wedge\cdots\wedge \diff x^{\mu_r}$, $\gamma_5\equiv \gamma_{1234}$.
The bilinears $S,P$ are scalars, $\xi^\flat, K$ are one-forms and $U$ is a two-form. These bilinears are in general all complex.
They satisfy various algebraic and differential relations implied by the Killing spinor equations, where the vector $\xi$, dual to 
$\xi^\flat$, is Killing. An important relation is 
\begin{align}\label{keyrel}
\dd \xi^\flat = & - \sqrt{2}\mskip2mu \ex^{\mathcal{K}/2}\xinew_I ( X^I U_{\selfd} - \widetilde{X}^I U_{\aselfd} ) \nonumber\\ & 
-\sqrt{2}\cI_{IJ} ( C^I F^{J}_{ \aselfd} -\widetilde{C}^I F^{J}_{ \selfd} )\,,
\end{align}
where the $[\pm]$ refer to self-dual and anti-self-dual parts, with respect to the volume form with $\epsilon_{1234}=+1$,
and we have defined
$C^I\equiv \ex^{\mathcal{K}/2}X^I(S-P)$, $\widetilde{C}^I\equiv \ex^{\mathcal{K}/2}\widetilde{X}^I(S+P)$.

The bilinears can be used to construct polyforms which are equivariantly closed under the action of 
$\dd_\xi\equiv \dd-\xi\hook\, $. Associated with the field strengths we have
\begin{equation}\label{eqvartFI}
\Phi_{(F)}^{I}=F^I+\Phi_0^I\,,\qquad \dd_\xi\Phi_{(F)}^{I}=0\,,
\end{equation}
where
\begin{align}\label{eqvartFI2}
\Phi_0^I\equiv \sqrt{2}\big(C^I-\widetilde{C}^I\big)\,.
\end{align}
We also have
\begin{equation}
    \Phi=\Phi_4+\Phi_2+\Phi_0\,, \qquad \dd_\xi\Phi=0\,,
\end{equation}
where 
\begin{align}\label{phi420}
\Phi_4&\equiv  -\frac{1}{2}\mathcal{V}\mskip2mu \mathrm{vol}_{4}-\frac{1}{4}\mathcal{I}_{IJ}F^I\wedge*F^J+\frac{\ii}{4}\mathcal{R}_{IJ}F^I\wedge F^J\,,\nn\\
\Phi_2
&\equiv    \frac{1}{\sqrt{2}}\ex^{\mathcal{K}/2}(WU_{\selfd}+\widetilde{W}U_{\aselfd})\nn\\ & 
\mkern-18mu -\frac{1}{\sqrt{2}}\mathcal{I}_{IJ}\big(C^IF^{J}_{ \selfd} 
 +\widetilde{C}^IF^{J}_{ \aselfd}\big)
+\frac{\ii}{\sqrt{2}}\mathcal{R}_{IJ}F^J\big(C^I-\widetilde{C}^I\big)\,,\nn\\
\Phi_0&\equiv   \ii\big[\mathcal{F}(C)-\mathcal{F}(\widetilde{C}) - \partial_I\mathcal{F}(C)\widetilde{C}^I+\partial_I\mathcal{F}(\widetilde{C})C^I\big]\,.
\end{align}
The existence of $\Phi_{(F)}^{I}$ allows us to compute various flux integrals using localization, while $\Phi$ allows us to compute the 
on-shell (bulk) action, $I_{\mathrm{OS}}$. Specifically, if we take the trace of the Einstein equation and substitute back into the action \eqref{eq:I4dEucl} we find 
\begin{align}\label{POSactdef}
I_{\mathrm{OS}}	&=\frac{\pi}{2G_4} \frac{1}{(2\pi)^2}\int_{M_4} \Phi_4\,.
\end{align}

The Killing spinor is charged with respect to the Lie derivative $\mathcal{L}_\xi$, with the charge dependent on the choice of gauge for $\zeta_IA^I$. Using a gauge with $\mathcal{L}_\xi A^I=0$, from~\eqref{eqvartFI} we have
$\xi\hook A^I=-\Phi_0^I+c^I$, where $c^I$ are constants, and one then computes
$\mathcal{L}_\xi\epsilon=\frac{\ii}{4}\zeta_I c^I\epsilon$.
Furthermore, at a fixed point, generalizing an argument in \cite{BenettiGenolini:2023ndb}, we have
 \begin{align}\label{wtsgam}
\frac{1}{8}\gamma^{\mu\nu}\diff \xi^\flat _{\mu\nu}\epsilon=\frac{\ii}{4}\zeta_I\Phi^I_0\mskip1mu \epsilon\,.
\end{align}

\section{Evaluating the action}
We are interested in solutions $M_4$ with a conformal boundary $\partial M_4= M_3$, and we allow that $M_4$ has certain
orbifold singularities on the fixed point nuts.
For simplicity, we now restrict to solutions with real metric and gauge fields. We also restrict to
solutions with $\tepsilon =\epsilon^c$, which implies that $z^i,\tz^{\tilde i}$ and $X^I,\widetilde X^I$ are real \footnote{The further subclass of solutions with $z^i=\tilde{z}^{\tilde{i}}$ are then associated with Lorentzian solutions.}. 
For this class of solutions the bilinears \eqref{eq:Bilinears} are all real. As we shall see, this class 
is still very rich.

To evaluate the on-shell gravitational action, we need to evaluate \eqref{POSactdef}, and then supplement it with boundary 
terms, associated with a supersymmetric renormalization scheme.
We assume that $\xi$ has no fixed points on the conformal boundary, which allows us to integrate \eqref{POSactdef} by parts, leading to a boundary term on $M_3=\partial M_4$ together with contributions around the fixed point set in the interior of $M_4$. 
Remarkably, this divergent boundary term exactly cancels with the other boundary contributions \cite{BenettiGenolini:2024lbj}; from a dual CFT perspective
this corresponds to a BPS condition on the energy in terms of the conserved charges. 
The on-shell action \eqref{POSactdef} then effectively receives contributions only 
from the fixed points, given by the 
BVAB formula \cite{BV:1982, Atiyah:1984px} 
\begin{align}\label{I4dexpand}
I_{\mathrm{OS}} = \frac{\pi}{2G_4}\bigg\{\sum_{\substack{\mathrm{nuts}}}\frac{1}{d}\frac{ \Phi_0}{b_1b_2} +
\sum_{\substack{\mathrm{bolts}} }\int_\Sigma 
\frac{\Phi_2}{2\pi b_2}-\frac{\Phi_0 c_1(L)}{b_2^2}\bigg\}\, .
\end{align}
Here $b_i$ are the weights of the linear action of $\xi$ on the normal bundle  to the fixed point set, $d$ is the order of the finite group
characterizing the orbifold singularity for a nut, and
$c_1(L)$ is the first Chern class of the normal bundle to the bolt surface $\Sigma$ in $M_4$.
  
Following \cite{BenettiGenolini:2019jdz}
we may write $P=S\cos\theta$, where  $\|\xi\|=
S \abs{\sin\theta}$. Since the spinor square norm $S$ is necessarily 
nowhere zero (see \cite{Ferrero:2021etw} for a general argument), 
it follows that at a fixed point $\cos\theta=\pm 1$ and hence
correspondingly $P=\pm S$,  so the Killing spinor $\epsilon$ is
necessarily chiral $\gamma_5\epsilon = \pm \epsilon$ at such a fixed locus. Notice then
that we have either $(C^I,\widetilde C^I)|_\pm=(0,2S\mskip1mu  \ex^{\mathcal{K}/2}\widetilde X^I)|_+$ or  $(2S \mskip1mu \ex^{\mathcal{K}/2}X^I,0)|_-$, respectively, and 
also 
$(\widetilde C^I/\zeta_J\widetilde C^J)|_+=u^I_+$ or $(C^I/\zeta_J C^J)|_-=u^I_-$, where $u^I_\pm$ were defined
in \eqref{uI}. 

For a nut with $\cos\theta=\pm1$, one can show that in a local
 orthonormal frame the Killing spinor satisfies  the projection conditions 
$\gamma^{12}\epsilon=- \ii\pmu  \mskip1mu \epsilon$, $\gamma^{34}\epsilon=\pm \ii\pmu  \mskip1mu \epsilon$. 
Here $\pmu=\pm 1$ can be set to $\pmu=1$ by an appropriate choice of conventions at a 
given fixed point, but when there is more than one connected component relative signs 
are important.
Equation \eqref{wtsgam} then immediately  leads 
to
\begin{align}\label{nutrelnswts}
(b_1-b_2)=\frac{\pmu}{\sqrt{2}}
\zeta_I\widetilde C^I\,,\quad
(b_1+b_2)=-
\frac{\pmu}{\sqrt{2}}
\zeta_I C^I\,,
\end{align}
for a nut$_\pm$, respectively. Using this in the expression for $\Phi_0$ in \eqref{phi420}, 
combined with the fact that the prepotential is homogeneous of degree two, 
one can evaluate the nut contribution in \eqref{I4dexpand} leading to the result for nuts in
\eqref{Fgrav}. 

To evaluate the contribution from bolts we proceed similarly. 
The local orthonormal frame is chosen so that the $\mathrm{e}^1$, 
$\mathrm{e}^2$ directions are tangent to the bolt. With the same 
projection conditions on the spinor as for a nut, with $b_2$  being 
the weight of $\xi$ on the normal $\R^2$ space to the bolt$_\pm$, 
equation  \eqref{wtsgam} gives
\begin{align}\label{boltbs}
 b_2=-\frac{\pmu }{\sqrt{2}}
\zeta_I\widetilde C^I\,,\qquad
 b_2=-
\frac{\pmu}{\sqrt{2}}
\zeta_I C^I\, ,
\end{align}
respectively. 
Focusing on the latter case,
the integral of $\dd\xi^\flat$ over the bolt vanishes using Stokes' theorem, and hence from \eqref{keyrel} we deduce that
$\int_{\Sigma_-}\ex^{\mathcal{K}/2}\zeta_I X^IU_{\selfd} = -\int_{\Sigma_-}\mathcal{I}_{IJ}C^IF^{J}_{\aselfd}$. 
This combined with
\eqref{phi420} implies $\int_{\Sigma_-}\Phi_2=\frac{\ii}{\sqrt{2}}\int_{\Sigma_-} C^I{\mathcal{N}}_{IJ}F^J$.
This can be further simplified: smoothness of the gauge field at the bolt implies that $\xi\hook F^I=0$ and then
\eqref{eqvartFI}, \eqref{eqvartFI2} imply that $C^I$ is constant on $\Sigma_-$ \footnote{In known solutions
the scalar fields and the bilinears $S,P$ are also constant on $\Sigma_-$ but we don't know whether this has to be the case in general.}; using
 $\mathcal{N}_{IJ}X^J=\mathcal{F}_I$, equation \eqref{boltbs} and the fact that $\mathcal{F}$ is homogeneous degree two then
allows us to show that the $\Phi_2$ contribution in \eqref{I4dexpand} for $\Sigma_-$ is as in~\eqref{Fgrav}. 
The $\Phi_0$ contribution is simple to evaluate. 
A similar computation for $\Sigma_+$ then completes the proof of~\eqref{Fgrav}.

\section{Flux integrals and UV-IR relation}

Consider integrals of $F^I$ over two-cycles in $M_4$. 
For $\Sigma\subset M_4$ a closed two-dimensional 
submanifold representing a cycle $[\Sigma]\in H_2(M_4,\Z)$, we may define the associated magnetic flux as \footnote{For gauged supergravity models that can be uplifted
to $D=10,11$ supergravity on compact manifolds, the magnetic fluxes are quantized. We have chosen the normalization here so that
$\mathfrak{p}^I\in\mathbb{Z}$ for the STU model.}
\begin{align}\label{pfrak}
\mathfrak{p}^I_{[\Sigma]} \equiv \frac{1}{2}\int_{\Sigma} c_1(F^I)=\frac{1}{4\pi}\int_{\Sigma} F^I\, . 
\end{align}
For a bolt $\Sigma=\Sigma_\pm$, these fluxes directly enter 
equation~\eqref{Fgrav}. If instead $\xi$ is tangent to $\Sigma$, 
hence rotating it, one can use localization 
to relate these fluxes to the fixed point data at nuts, as we 
illustrate in the examples below.

We can also obtain important  ``UV-IR relations'' by considering integrals over non-compact two-dimensional submanifolds, whose boundaries end on 
$\partial M_4= M_3$. We assume the submanifolds have
topology $\mathbb{R}^2$, with a boundary circle $S^1_{\mathrm{UV}}\equiv \partial\R^2\subset M_3$, and 
are invariant under the action of $\xi$. On such an $\mathbb{R}^2$ we may write $\xi\mskip1mu |_{\R^2}=b_2\partial_{\varphi_2}$, with $\Delta\varphi_2=2\pi$, and we refer to
the origin of $\mathbb{R}^2$, lying on the fixed point set, as the ``IR''. We define
\begin{align}
 \Delta^I\equiv \frac{1}{4\pi}\int_{\R^2} F^I\,,
\end{align}
which is gauge-invariant. Using Stokes' theorem we have  $\Delta^I
 =\frac{1}{4\pi}( \int_{S^1_{\mathrm{UV}}} A^I - \int_{S^1_{\mathrm{IR}}} A^I)$, where $S^1_{\mathrm{IR}}$ is a small circle surrounding the origin
 of $\mathbb{R}^2$. Choosing a gauge where $A^I$ is a global one-form on $\mathbb{R}^2$, the IR integral vanishes and $ \Delta^I$ fixes the holonomy of the
gauge field in the boundary theory. 
We next define
\begin{align}
\sigma^I  \equiv -\frac{\ii }{4\pi}\Phi^I_0\mskip1mu |_{\mathrm{UV}}\, ,\qquad
y^I \equiv \frac{ 1}{2b_2}\Phi^I_0 \mskip1mu |_{\mathrm{IR}}\, ,
\end{align}
with $\sigma^I$ characterizing UV scalar deformations in the dual boundary theory. 
From the equivariant closure of  $\Phi^{I}_{(F)}$ one then immediately deduces
\begin{align}\label{UVIR}
\Delta^I +\ii  \beta\sigma^I = y^I\, ,
\end{align}
where we write $\xi=\partial_\psi$ with $\Delta\psi =2\pi/b_2\equiv \beta$.  

We may relate the IR fixed point quantities $y^I$ to the fixed 
point data $u^I_\pm$ entering \eqref{Fgrav} as follows.
On the tangent space to the origin of the above $\R^2\subset M_4$, 
 we can write  $\xi=b_1\partial_{\varphi_1}+b_2
\partial_{\varphi_2}$, where $\varphi_1$ is a polar coordinate 
on the $\R^2$ that is normal to the $\R^2$ considered above, i.e. $\R^4=\R^2\oplus\R^2$ 
at the fixed point, with $\partial_{\varphi_i}$ rotating each factor. The case $b_1=0$ is then a bolt, while $b_1\neq 0$ is a nut. 
For a fixed bolt, from \eqref{boltbs} 
we have 
$y^I_\pm= \pm \pmu  \mskip1mu u^I_\pm$. For a fixed nut 
\eqref{nutrelnswts} implies 
\begin{align}\label{yureln}
y^I_\pm= \pm \pmu \mskip1mu (1\mp \omega) u^I_\pm\,, \quad \mbox{where} \ \omega\equiv \frac{b_1}{b_2}\, .
\end{align}
Notice this expression is valid both when the origin of the
$\mathbb{R}^2$ is a nut$_\pm$ or a bolt$_\pm$.
Substituting into
\eqref{UVIR} then gives a ``UV-IR relation'', valid for any $\xi$-invariant submanifold $\mathbb{R}^2$ that
has a UV boundary $S^1_\mathrm{UV}$.
A corollary is that for all such $S^1_\mathrm{UV}$
the UV data is constrained via
 \begin{align}
 \zeta_I (\Delta^I +\ii  \beta\sigma^I )= \zeta_I y_\pm^I =\pm \pmu \mskip1mu { (1\mp \omega)}\,.
 \end{align}
 Additional constraints can also be obtained from the magnetic fluxes,  as we illustrate in the examples below.

\section{Examples}
We can immediately
recover the result for minimal supergravity
\cite{BenettiGenolini:2019jdz} from \eqref{Fgrav}. 
It is also
straightforward to recover other results 
in the literature, and also obtain new formulae. We illustrate with 
the STU model, for which  
\begin{align}
\cF(X^I)
= -2\ii \sqrt{X^0X^1X^2X^3}\, ,
\end{align}
with $\xinew_I=1$ for all $I=0,1,2,3$.
Solutions to this theory uplift on $S^7$ to solutions 
of $D=11$~supergravity \cite{Cvetic:1999xp} and in our conventions
$\frac{\pi}{2G_4}=F^{\mathrm{ABJM}}_{S^3}$ is the free energy of the dual ABJM theory on $S^3$ in the large $N$ limit.

\subsection{Deformations of Euclidean \texorpdfstring{$AdS_4$}{AdS4}}

Let $M_4$ topologically be $\R^4 = \R_1^2\oplus \R_2^2$, with conformal 
boundary $\partial M_4=M_3\cong S^3$.
We take 
$\xi = \sum_{i=1}^2 b_i\partial_{\varphi_i}$ where $\partial_{\varphi_i}$ rotate 
each $\R_i^2$. We do not make any further assumptions on the metric on
$M_4$ or the boundary $S^3$.
For $b_i\neq 0$ there is an isolated fixed point at the origin of $\R^4$, 
and from \eqref{Fgrav} we may write 
\begin{align}\label{FP}
\Fgrav = \mp \frac{4(b_1\mp b_2)^2}{b_1b_2}\sqrt{u^0_\pm u^1_\pm u^2_\pm u^3_\pm}
F^{\mathrm{ABJM}}_{S^3}\, .
\end{align}
Here the $\pm$ signs are correlated with the chirality of the spinor at the origin, 
and from \eqref{uI} we have the constraint $\sum_{I=0}^3 u^I_\pm = 1$. 

This incorporates two distinct solutions in the literature: 
(i) taking $S^3$ to have a round metric and setting $b_1=\mp b_2$ gives the $SO(4)$-symmetric solutions 
of \cite{Freedman:2013oja} (for real $z^i,\tz^{\tilde i}$), where the two choices of sign 
were referred to as different ``branches'' of solution. For $b_1=\mp b_2$
the R-symmetry is embedded in $U(1)_\mp\subset SU(2)_\mp$, where 
$SU(2)_+\times SU(2)_-$ is the double cover of the $SO(4)$ 
isometry. 
In this case $\Delta^I=0$, and the UV-IR relation \eqref{UVIR} 
relates $u^I = 2\pi \ii \sigma^I$, where $\sigma^I$ are defined on
the $S^3$ UV boundary. {This case is dual to ABJM theory on $S^3$ with real mass 
deformations, and matches the large $N$ limit of the partition function \cite{Jafferis:2011zi}.}
The $AdS_4$ vacuum is recovered by setting $u^I_\pm =\tfrac{1}{4}$ for all $I$, which also extremizes 
\eqref{FP} as a (constrained) function of $u^I_\pm$; (ii) 
instead setting $u^I_\pm=\tfrac{1}{4}$ for all $I$ but keeping $b_i$ general, the $U(1)^2$-invariant metric on $S^3$ is now 
squashed, reproducing the minimal supergravity solutions of 
 \cite{Martelli:2011fu,Farquet:2014kma}.

The supergravity solutions with $M_4\cong\R^4$ with generic $b_i$ and $u^I_\pm$ have not been 
constructed, but assuming they exist  \eqref{FP}  gives their free energy.
Generically they will have $U(1)^2$ isometry, and thus be cohomogeneity two, with a squashed metric on the boundary $S^3$.
The result in \eqref{FP} is in precise agreement with the large $N$ result in \cite{Bobev:2022eus}.

\subsection{Black saddle 
solutions} 
Euclidean, dyonically charged, ``black saddle'' solutions of the STU model were 
analysed in \cite{Bobev:2020pjk}, with topology $M_4\cong\R^2\times \Sigma_g$, where 
$\Sigma_g$ is a genus $g$ Riemann surface. 
The special subclass of solutions with $\widetilde{X}^I=\overline{X}^I$ can be Wick rotated 
to Lorentzian signature extremal black holes with horizon $\Sigma_g$. 
The R-symmetry Killing vector rotates the Euclidean time circle, which 
is the polar direction in $\R^2$, and thus the horizon 
is a fixed bolt. 
The normal bundle is trivial, so $c_1(L)=0$ and 
from \eqref{Fgrav} (with $\pmu=1$) we have 
\begin{align}\label{BS}
\Fgrav= -2\sqrt{u^0_\pm u^1_\pm u^2_\pm u^3_\pm}\sum_{I=0}^3\frac{\mathfrak{p}^I}{u^I_\pm}F^{\mathrm{ABJM}}_{S^3}
\, ,
\end{align}
where 
$\mathfrak{p}^I=
\int_{\Sigma_g}F^I/4\pi$ as in \eqref{pfrak}, which by \eqref{RFlux} are constrained to satisfy $\sum_{I=0}^3 \mathfrak{p}^I = - 2(1-g)$. The 
scalars  $u^I_\pm$ are necessarily constant on the horizon $\Sigma_g$, 
and are related to the UV variables $\Delta^I+\ii\beta \sigma^I$  on the conformal 
boundary $M_3=S^1\times \Sigma_g$
via the general UV-IR relation \eqref{UVIR}, \eqref{yureln}, thus proving
the conjecture of \cite{Bobev:2020pjk} (who verified this relation
for a subset of analytic and numerical solutions).
The result \eqref{BS} agrees with a large $N$ limit 
of the topologically twisted index of the ABJM theory 
on  $M_3=S^1\times \Sigma_g$ \cite{Bobev:2020pjk} (for $g\ne 1$).

\subsection{Taub-bolt saddle solutions}
One can generalize the black saddle construction of \cite{Bobev:2020pjk} by considering
the boundary $M_3$ to have the $S^1$ fibred over $\Sigma_g$, 
with Chern number $-p\in\Z$. The large $N$ limit 
of the partition function of the ABJM theory on this space was computed in  \cite{Toldo:2017qsh}. 
A natural candidate gravity dual has $M_4=\mathcal{O}(-p)
\rightarrow\Sigma_g$, the total space of the associated 
complex line bundle over $\Sigma_g$. Such solutions 
to the Euclidean STU theory are not known, although 
there are explicit solutions in minimal supergravity 
constructed in \cite{Martelli:2012sz,Toldo:2017qsh}.
Assuming these solutions exist, 
they have a bolt with non-trivial normal bundle and
$\int_{\Sigma_g}c_1(L)=-p$. Thus, from \eqref{Fgrav} ({with $\pmu=1$}) we have 
\begin{align}\label{TW}
\Fgrav= -2 \sqrt{u^0_\pm u^1_\pm u^2_\pm u^3_\pm}
\Big( \pm 2p+ \sum_{I=0}^3\frac{\mathfrak{p}^I_\pm}{u^I_\pm}\Big)
F^{\mathrm{ABJM}}_{S^3}\, ,
\end{align}
with $\sum_{I=0}^3 \mathfrak{p}^I_\pm =\mp p - 2(1-g)$ from \eqref{RFlux} and
with the same UV-IR relation as in the previous subsection. 
Remarkably, \eqref{TW} is in
precise agreement with the 
large $N$ field
 theory result of \cite{Toldo:2017qsh}.

\subsection{Spindle solutions}

Now consider solutions with topology $M_4=\R^2\times \Sigma$, 
where $\Sigma\cong\mathbb{WCP}^1_{[n_N,n_S]}$ is a 
spindle. The latter is topologically a
two-sphere, but with conical deficit angles $2\pi(1 - 1/n_{N,S})$
at the north and south poles, 
respectively. 
These correspond to dyonic black saddle solutions with acceleration 
and rotation; 
such solutions in minimal gauged supergravity 
have been constructed \cite{Ferrero:2020twa}. 

We parametrize
$\xi =  \varepsilon \mskip1mu \partial_{\varphi_1}+ b \mskip1mu \partial_{\varphi_2}$, 
where $\partial_{\varphi_1}$, $\partial_{\varphi_2}$ rotate 
the  spindle and $\R^2$ factors, respectively, 
and $\Delta\varphi_1=2\pi=\Delta\varphi_2$. 
We choose sign conventions so that 
the fixed north pole $N$ has negative chirality, 
while the south pole $S$ has chirality $-\sigma=\mp 1$, 
which are the twist and anti-twist of \cite{Ferrero:2021etw}. 
The weights at the poles are $(\varepsilon_N/n_N,b_N)\equiv(-\varepsilon/n_N,1)$ and $(\varepsilon_S/n_S,b_S)\equiv(\varepsilon/n_S,1)$.  
Applying \eqref{Fgrav} with $d_N=n_N$, $d_S=n_S$ immediately 
gives 
a ``gravitational block" \cite{Hosseini:2019iad} type of expression:
\begin{align}
\Fgrav 
& = -\frac{4}{\varepsilon}\Big[\sqrt{y_N^0y_N^1y_N^2y_N^3}
-\sigma \sqrt{y_S^0y_S^1y_S^2y_S^3}\mskip3mu\Big]F^{\mathrm{ABJM}}_{S^3}\, ,
\end{align}
where the $y^I_{N,S}$ variables satisfy the constraints
\begin{align} 
\sum_{I=0}^3 y^I_N = \frac{\varepsilon}{n_N}-1
\,, \quad 
\sum_{I=0}^3 y^I_S = -\sigma\frac{\varepsilon}{n_S}-1\, ,
\end{align}
where we have taken $\pmu_N=1$, $\pmu_S=\sigma$.
The fluxes can be obtained by localizing $\Phi^I_{(F)}$, leading to
\begin{align}\label{eq:fluxspindle}
	\mathfrak{p}^I=\frac{2}{\varepsilon}\left(\left.\Phi_0^I\right|_N-\left.\Phi_0^I\right|_S\right)=-\frac{1}{\varepsilon}(y_N^I-y_S^I)\,, 
\end{align}
which implies $\sum_{I=0}^3 \mathfrak{p}^I=-\frac{n_S+\sigma n_N}{n_N n_S}$.
These results agree with \cite{BenettiGenolini:2024kyy} after
identifying $y^I_{N,S}= -x^I_{\pm}/2$, $\varepsilon=b_0$. 

\section{Discussion}\label{sec:discussion}
Our derivation of \eqref{Fgrav} assumed solutions with real fields, but we expect this
can be relaxed. For example, our results are in agreement with the solutions in \cite{Freedman:2013oja} both for real and complex scalars.
Furthermore, the on-shell action that we obtained for putative Euclidean spindle solutions also gives rise to the on-shell action
for a complex locus of accelerating black hole solutions, and hence the entropy (after a Legendre transform) \cite{Cassani:2021dwa,Ferrero:2021ovq}.
 
Other generalizations include adding hypermultiplets, which for some models will be associated with a Higgs mechanism in the bulk,
 leading to additional constraints on the parameters $u_\pm^I$, similar to \cite{BenettiGenolini:2024kyy}, and generalizing
to $D=5$. We also anticipate that we can incorporate higher derivatives, 
and this will provide a promising way to prove and extend the conjectures discussed in \cite{Hristov:2021qsw,Hristov:2024cgj}
as well as make contact with $1/N$ corrections to field theory results as in e.g. \cite{Bobev:2022eus}.

\section*{Acknowledgments}
This work was supported by STFC grants  ST/T000791/1 and
ST/T000864/1, EPSRC grant EP/R014604/1 and SNSF Ambizione grant PZ00P2\_208666.
JPG is a Visiting Fellow at the Perimeter Institute. 
AL is supported by a Palmer Scholarship.

\bibliography{biblio}

\bigskip

\appendix

\section{Supplementary material}

The supersymmetry transformations of the Euclidean theory
are parametrized by two Dirac spinors $\epsilon$ and $\tilde \epsilon$, with corresponding Killing spinor equations  given by  
\begin{align}
\label{eq:Euclidean_KSE_epsilon}
	0 &= D_\mu \epsilon
+ \frac{1}{2\sqrt{2}}  \gamma_\mu  \e^{\cK/2}\Big( W \newP_-  + \widetilde{W} \newP_+ \Big)\epsilon \nn\\
& - \frac{\ii}{4\sqrt{2}} \mathcal{I}_{IJ} F^{J}_{\nu\rho} \gamma^{\nu\rho}\gamma_\mu \ex^{\mathcal{K}/2}\left( X^I \newP_- + \widetilde{X}^I \newP_+ \right) \epsilon \, ,\\
	0 &= \frac{\ii}{2\sqrt{2}} \mathcal{I}_{IJ} F^{J}_{\nu\rho} \gamma^{\nu\rho} \e^{\cK/2}\left( \cG^{\tilde{i} j} \nabla_j X^I \newP_- + \cG^{i\tilde{j}} \nabla_{\tilde j} \widetilde{X}^I \newP_+ \right) \epsilon\nonumber\\ 
	&\ \ \ + \gamma^\mu \left( \partial_\mu z^i \newP_- + \partial_\mu \tilde{z}^{\tilde{i}} \newP_+ \right) \epsilon \nonumber\\
	& \ \ \  -  \frac{1}{\sqrt{2}}  \ex^{\mathcal{K}/2} \left( \cG^{\tilde{i} j} \nabla_j W \newP_- +  \cG^{i\tilde{j}} \nabla_{\tilde j} \widetilde{W} \newP_+ \right) \epsilon \, , \\
\label{eq:Euclidean_KSE_tepsilon}
	0 &= \widetilde{D}_\mu \tepsilon  
+ \frac{1}{2\sqrt{2}}  \gamma_\mu \Big( W \newP_- + \widetilde{W} \newP_+ \Big) \tepsilon \nonumber\\
	& \ \ \ + \frac{\ii}{4\sqrt{2}} \mathcal{I}_{IJ} F^{J}_{\nu\rho} \gamma^{\nu\rho}\gamma_\mu \ex^{\mathcal{K}/2}\left( X^I \newP_- + \widetilde{X}^I \newP_+ \right) \tepsilon \, ,\\
	0 &= - \frac{\ii}{2\sqrt{2}} \mathcal{I}_{IJ} F^{J}_{\nu\rho} \gamma^{\nu\rho}  \e^{\cK/2}\left( \cG^{\tilde{i} j} \nabla_j X^I \newP_- + \cG^{i\tilde{j}}  \nabla_{\tilde j} \widetilde{X}^I \newP_+ \right) \tepsilon \nonumber\\
	& \ \ \ + \gamma^\mu \left( \partial_\mu z^i \newP_- + \partial_\mu \tilde{z}^{\tilde{i}} \newP_+ \right) \tepsilon \nonumber\\
	& \ \ \  -  \frac{1}{\sqrt{2}} \ex^{\mathcal{K}/2} \left( \cG^{\tilde{i} j} \nabla_j W \newP_- + \cG^{i\tilde{j}} \nabla_{\tilde j} \widetilde{W} \newP_+ \right) \tepsilon \, .
\end{align}
Here
$\newP_\pm \equiv \frac{1}{2}(1 \pm \gamma_5 )$, 
with covariant derivatives 
$D_\mu\epsilon \equiv \nabla_\mu \epsilon + \frac{\ii}{2} \cA_\mu \gamma_5 \epsilon -  \frac{\ii}{4} \zeta_I A_\mu^I  \epsilon$, 
$\widetilde{D}_\mu\tepsilon\equiv\nabla_\mu \tepsilon + \frac{\ii}{2} \cA_\mu \gamma_5 \tepsilon +  \frac{\ii}{4} \zeta_I A_\mu^I  \tepsilon$.

\end{document}